\documentclass[conference]{IEEEtran}
\IEEEoverridecommandlockouts
\usepackage{cite}
\usepackage{amsmath,amssymb,amsfonts}
\usepackage{algorithmic}
\usepackage{graphicx}
\usepackage{textcomp}
\usepackage{xcolor}
\usepackage{tabularx}
\usepackage{makecell}
\usepackage[bookmarks=false]{hyperref}
\usepackage{enumitem}
\usepackage{ulem}
\usepackage{xcolor}

\usepackage[font=small, labelfont=sc, labelsep=colon]{caption}

\def\BibTeX{{\rm B\kern-.05em{\sc i\kern-.025em b}\kern-.08em
    T\kern-.1667em\lower.7ex\hbox{E}\kern-.125emX}}
\begin{document}

\title{Real-World Modeling of Computation Offloading for Neural Networks with Early Exits and Splits\\
	\thanks{ This work was supported by Ministry of Education, Youth and Sport of the Czech Republic under grant no. LUASK22064.
	
	This work has been submitted to IEEE GLOBECOM 2025 for peer review.}
	
}

\author{\IEEEauthorblockN{Jan Danek, Zdenek Becvar, and Adam Janes}
	\IEEEauthorblockA{Faculty of Electrical Engineering, Czech Technical University in Prague, Prague, Czech Republic}
	
	\IEEEauthorblockA{danekja5@fel.cvut.cz, zdenek.becvar@fel.cvut.cz, janesada@fel.cvut.cz}}

\maketitle

\begin{abstract}
We focus on computation offloading of applications based on convolutional neural network (CNN) from moving devices, such as mobile robots or autonomous vehicles, to Multi-Access Edge Computing (MEC) servers via a mobile network. In order to reduce overall CNN inference time, we design and implement CNN with early exits and splits, allowing a flexible partial or full offloading of CNN inference. Through real-world experiments, we analyze an impact of the CNN inference offloading on the total CNN processing delay, energy consumption, and classification accuracy in a practical road sign recognition task. The results confirm that offloading of CNN with early exits and splits can significantly reduce both total processing delay and energy consumption compared to full local processing while not impairing classification accuracy. Based on the results of real-world experiments, we derive practical models for energy consumption and total processing delay related to offloading of CNN with early exits and splits.
\end{abstract}


\section{Introduction}
Autonomous systems%TODO\textcolor{red}{pokud je cely clanek o AV, tak pak abych zkratku AS ani nezavadel a pouzival cely nazev, at to neni matouci; nebo jeste lepe nezavadet zkratku AV a mluvit obecne o AS a tam kde je to nutne specifikovat, tak uvest autonomous vehicle Al eje treba v clanku (opakovane) zruraznit, ze reseni je pro AV i (mobile) robots}
, such as autonomous vehicles (AVs) or robots, rely on real-time data processing to operate safely and effectively in dynamic environments \cite{Karangwa2023}. However, onboard computing resources of autonomous systems are often limited and may be insufficient for demanding data processing tasks (for example, computer vision). To address these limitations, multi-access edge computing (MEC) has emerged as a promising approach. MEC enables a partial or full offloading of computationally intensive tasks from the autonomous systems to nearby computing servers located close to base stations (BSs) in mobile networks, thereby reducing the total task processing delay, including both communication and computing delays \cite{Liu2022}. 

%\cite{Becvar2017}

Many recent works optimize utilization of the MEC resources via allocation of communication and computing resources to maximize the number of processed tasks \cite{Liu2022},
%TODO \textcolor{ForestGreen}{ \cite{Li2022}}. 
While the number of processed tasks is minimized, the total processing delay is not considered in such works despite the delay is essential metric to ensure real-time operation of the autonomous systems. In addition, existing works, such as \cite{Hu2022} %TODO\textcolor{ForestGreen}{and \cite{Sheng2020}} 
typically focus on uplink transmission during CNN offloading to MEC server and neglect downlink delivery of the computing  results. While in simplified theoretical communication models such assumption may be applicable, in practical application, downlink results delivery back to the autonomous system is not negligible despite a low amount of data due to overhead imposed by frame alignment, scheduling, processing latency, and buffering delays, as shown in \cite{Danek2025}. 

Furthermore, resource allocation for computation offloading depends on type of applications being processed. 
Due to the explosion of applications related to computer vision in both theory and practice, many recent works deal with offloading of machine learning-based computing tasks. An example of such task is convolutional neural network (CNN) for image processing (detection of objects in an image). Such type of applications, widely used in the AVs, can benefit from offloading from the AV to the MEC server due to high computational and energy demands of the CNN inference \cite{LiN2022}.
To reduce computing burden related to CNN inference, novel CNN architectures support early exits \cite{Casale2024}. The CNN architecture with early exits allow to classify inputs (e.g., objects in images) at earlier stages of the CNN provided that a required prediction confidence is reached \cite{Teerapittayanon216}, %\textcolor{ForestGreen}{, \cite{Huang2024}, \cite{Wang2024}}, 
thus reducing the overall CNN processing delay. 
%The paper \cite{Teerapittayanon216} introduces a pioneering framework that leverages early exit to accelerate inference in CNN with many layers. \textcolor{ForestGreen}{Follow-up works, such as \cite{Huang2024} analyzed the robustness of early exit CNN architecture, while \cite{Wang2024} proposed a class-exclusion mechanism to improve the overall CNN processing delay and robustness of the CNN in early exit CNN.}

Another option to reduce the delay of the CNN task processing is to split the CNN processing between the AV and the MEC server so that a part of layers is processed locally by the AV and a part is offloaded to the MEC server \cite{Bakhtiarnia2024}, \cite{Zhou2023}. The offloading of a part of CNN inference to the MEC server takes place at predefined split points. The split of CNN allows to accelerate the CNN inference in scenarios with the AVs that have limited computing power. The paper \cite{Eshratifar2021} presents a system that determines which layers of the CNN should be computed locally or remotely to minimize total processing delay and total energy consumption. 
%TODO \textcolor{ForestGreen}{Then, in \cite{Tuli2023}, the authors explore a selection of the optimal split point in a large CNN to improve processing efficiency in the MEC server.}

Another step beyond stand-alone early exists and stand-alone split is a combination of both in single CNN. Recent papers  \cite{Narmeen2024} and \cite{Rauch2024} explore the combination of early exits and split computing to optimize CNN inference with focus on reducing total processing delay. While all the works on early exits, splits, and their combination demonstrate notable improvement in various metrics of CNN-based task processing in the mobile networks with MEC, the performance is evaluated only by simulations under often over-optimistic assumptions. As a result, all existing works do not capture key aspects of practical deployment, such as hardware constraints or variations and processing in mobile networks.

%\cite{Bajpai2024}, \cite{Pacheco2024},

Motivated by lack of real-world evaluation and simplified assumptions and models in existing works, we implement and evaluate computing offloading from the AVs to the MEC servers for CNN-based applications with split points and early exits on real hardware considering all practical aspects, such as often neglected downlink communication delay, and we demonstrate the feasibility and performance trade-offs in the real-world environment. We focus on road sign classification using a resource-constrained AV, where CNN inference is either processed locally or (partially) offloaded to the MEC server over the mobile network. The major contributions of this paper are as follows:

\begin{enumerate}
	\item We implement offloading of CNN with early exits and split points from the AV to the MEC server. The implementation is done with real AV hardware including communication and computing components and MEC server to demonstrate feasibility of early exists and splits in practical deployment scenarios. Unlike related works, we consider both uplink and downlink communication, providing a comprehensive and realistic analysis of the communication delay. We also consider complete model of computing including preprocessing of data for computing and preparation of data to for transmission to the BS and to the MEC server. 
	\item We demonstrate the impact of CNN's early exits and splits on total computing task processing delay, energy consumption of the AV, and accuracy of CNN decisions. 
	\item Based on the experiments, we derive mathematical models for the total processing delay and total energy consumption for the applications based on CNN with early exits and splits to allow future theoretical research to be done with realistic models.
	\item All developed codes for all aspects of the autonomous systems and CNN, as well as experimental data are open source and publicly available on GitLab\footnote{\href{gitlab.fel.cvut.cz/mobile-and-wireless/codes/multi-exit-neural-network}{gitlab.fel.cvut.cz/mobile-and-wireless/codes/multi-exit-neural-network}}  for future research contributing to transparency and reproducibility of the work.
\end{enumerate}

The rest of the paper is organized as follows. First, we outline a model of the system considered in this paper. Then, in Section \ref{Implementation}, we describe our implementation of CNN with early exits and split points and integration of CNN to the testbed composed of MEC server, AV, and mobile network for offloading. Then, the evaluation scenario for experiments is outlined in Section \ref{ExperimentalSetup}. Section \ref{Evaluation} presents and discusses results of real-world experiments. In Section \ref{Modeling}, we introduce the analytical models of the total processing delay and total energy consumption.  Last, Section \ref{Conclusion} concludes the paper.

\section{System model} \label{systemModel}
In this section, we introduce modelling of the computation offloading of CNN from the AV to the MEC server via mobile network. We also define classification accuracy, the total processing delay and total energy consumption related to both offloading and local computing.

\begin{figure}[t]
	\centerline{\includegraphics[width=0.8\linewidth]{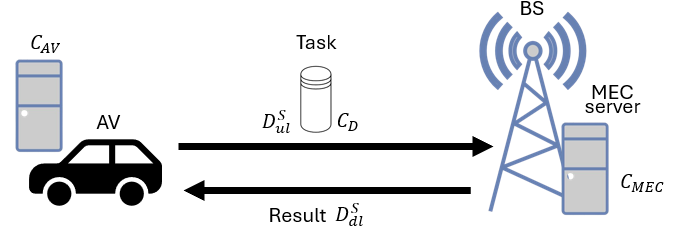}}
	\caption{System model with AV either computing/processing CNN interference locally or offloading the computation to MEC server via mobile network.}
	\label{Fig:sysetmModelAV}
\end{figure}

We consider a system composed of a single AV and a single BS, as illustrated in \ref{Fig:sysetmModelAV}. The BS is enhanced with the MEC server for the processing of computing tasks. In this work, we focus on the tasks based on machine learning, such as computer vision tasks including road sign %detection and 
classification, based on CNNs.

The CNN model used in our system, see Fig.~\ref{Fig:sysetmModelAV}, consists of $N_L$ layers grouped into $N_B$ blocks, where each block comprises a set of layers designed to extract specific features. The CNN includes $N_E$ early exits \cite{Casale2024} and $N_S$ split points \cite{Bakhtiarnia2024}. The split points are placed between blocks to reflect the CNN’s structure \cite{Matsubara2020}, and early exits are co-located with split points to simplify design by sharing intermediate outputs.

\begin{figure}[t]
	\centerline{\includegraphics[width=0.8\linewidth]{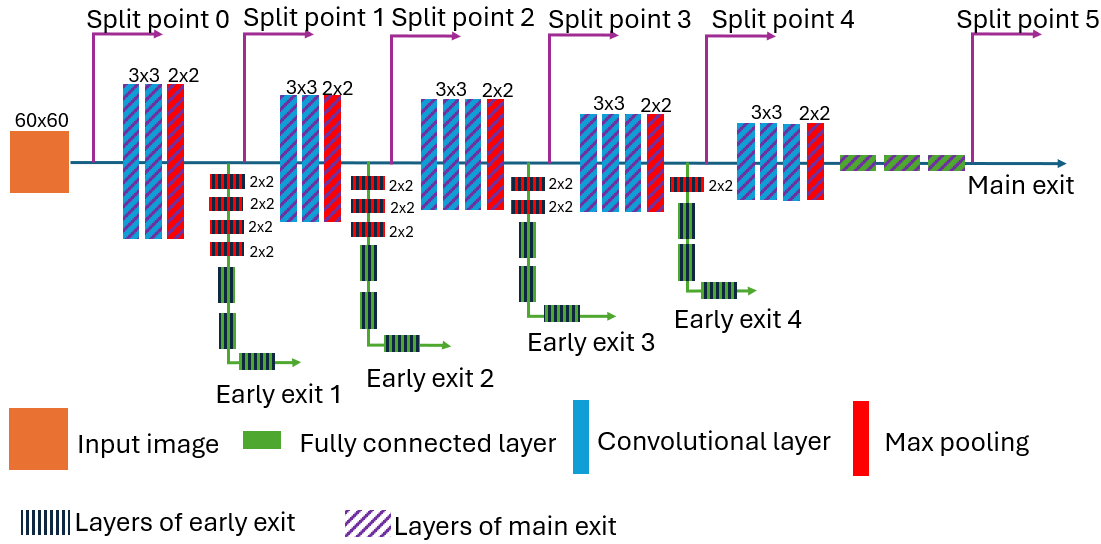}}
	\caption{Model of early exit CNN with several splits.}
	\label{Fig:neuralNetwork}
\end{figure}

The CNN can be processed fully on the AV, fully offloaded to the MEC server, or partially offloaded at any selected split point so that the layers before the split point are processed by the AV and remaining layers are processed by the MEC server. We define the variable $S$, which determines position of the split point within the CNN model, see Fig.~\ref{Fig:neuralNetwork}. All layers before this split point are processed locally by the AV while all layers after split point are processed by the MEC server. If $S=0$, all layers are processed by the MEC server. Conversely, if $S=N_S$, all layers are processed locally by the AV. 

We also define $E$ as the selected exit point of the CNN. If $E=N_E$, the main exit is used while $E<N_E$ represents the used early exit.
The selection of the early exits influences the trade-off between computing cost and classification accuracy $A(E,S)$, which is defined as:
\begin{equation}
	A(E,S) = \frac{N_{true}(E,S)}{N},
	\label{classification_accuracy}
\end{equation}
where $N_{true}(E,S)$ represents the number of correctly classified objects, and $N$ is the total number of inputs to the CNN. 

To optimize data transmission efficiency between the AV and the MEC server, an autoencoder is deployed at each split point $S$ to compress the output of CNN layers before transmission \cite{Matsubar2023}. Without compression, the data size can even exceed volume of the original input image, increasing uplink communication delay and total energy consumption \cite{Xu2003}. The encoder (on the AV) and the decoder (on the MEC) are implemented as convolutional layers, allowing continued CNN inference on reconstructed data \cite{Rauch2024}.

The compression ratio $R_S$ quantifies data reduction at the split point $S$ and is defined as the ratio of the original data volume $D_{orig}^S$ to the compressed data volume $D_{comp}^S$:

\begin{equation} 
	R_{S}=\frac{D^{S}_{orig}}{D^{S}_{comp}}. 
	\label{compress_ratio} 
\end{equation}

To enable partial offloading, CNN inference is divided into subtasks between consecutive split points \cite{Rauch2024}. Each subtask is characterized by:

\begin{itemize}
	\item Volume of data $D_{ul}^S$ transferred from the AV to the MEC server. The transmitted volume depends on the selected split point $S$ and equals $D_{ul}^S=0$ in case of local processing.
	\item Volume of computing results $D_{dl}^S$ sent back from the MEC server to the AV. This volume remains the same for all split points except local processing of the entire CNN, in this case  $D_{dl}^S=0$.
	\item Computing demand $C_D=(C^1_D,…,C^{N_S}_D)$, where each $C^i_D$ represents the required computing demand of the CNN layers between two consecutive split points.
\end{itemize}

The computing power of the AV and the MEC server is denoted as $C_{AV}$ and  $C_{MEC}$, respectively. Computation at the AV is done at a central processing unit (CPU) with a computing power $C_{AV}$. The MEC server is equipped with the CPU with a computing power $C_{CPU}$ and with a graphical processing unit (GPU) with a computing power $C_{GPU}$. The total computing power of the MEC server is  $C_{MEC}=C_{CPU}+C_{GPU}$.

The communication between the AV and the BS is characterized by an uplink bitrate $b_{ul}$ and a downlink bitrate $b_{dl}$, both depending on modulation, coding rate, and number of allocated resource blocks \cite{3GPP2022}. Thus, the uplink bitrate $b_{ul}$ is defined as:

\begin{equation}
	b_{ul} = N^{RB}_{ul} \times N^{sub}_{ul} \times N^{bits}_{ul} \times N^{sym}_{ul} \times CR_{ul},
	\label{uplink_bitrate}
\end{equation}
where $N_{ul}^{RB}$ is the number of used resource blocks, $N_{ul}^{sub}$ is the number of subcarriers per resource block, $N_{ul}^{bits}$ is the number of bits per symbol, $N_{ul}^{sym}$ is the number of symbols per subcarrier, and $CR_{ul}$  is the code rate. The downlink bitrate $b_{dl}$ is defined analogously with corresponding downlink parameters. %The bitrate $b_{dl}$ in downlink is defined analogically as for the uplink, just the uplink parameters $N_{ul}^{RB}$, $N_{ul}^{sub}$, $N_{ul}^{bits}$, $N_{ul}^{sym}$, $CR_{ul}$ are substituted with the related downlink parameters $N_{dl}^{RB}$, $N_{dl}^{sub}$, $N_{dl}^{bits}$, $N_{dl}^{sym}$, $CR_{dl}$, respectively.

The overall communication delay $t_n^{comm,S}$  for the $n$-th CNN interference includes the uplink communication delay $t_n^{ul,S}=d_{ul}+\frac{D_{ul}^S}{b_{ul}}$ for offloading the computing task to the MEC server and the downlink communication delay $t_n^{dl,S}=d_{dl}+\frac{D_{dl}^S}{b_{dl}}$ for transferring the result back to the AV. The uplink delay $d_{ul}$ and the downlink delay $d_{dl}$ denote constant processing delays caused by mobile network processing overhead at the UE and BS, respectively. The processing delay arises from factors such as frame alignment, scheduling, processing latency, and buffering delays \cite{Danek2025}.

The overall communication delay of the $n$-th CNN interference is formulated as:

\begin{equation}
	t_n^{\text{comm},S} =
	\begin{cases}
		t_n^{ul,S}  + t_n^{dl,S} =
		{}\\
		\left( d_{ul} + \frac{D_{ul}^S}{b_{ul}} \right) + \left( d_{dl} + \frac{D_{dl}^S}{b_{dl}} \right), & \text{if split } S \text{ is used}, \\
		0, & \text{otherwise}.
	\end{cases}
	\label{communication_delay}
\end{equation}

The computing delay $t_n^{comp}(E,S)$ of the $n$-th CNN interference is determined as:

\begin{equation}
	\begin{split}
		&t^{comp}_{n}(E,S) = t^{AV}_{n}(E,S)+t^{MEC}_{n}(E,S)\\
		 &= \left(t^{prep,S}_n + \sum_{i=1}^{\text{min}(S,E)}\frac{C_{D}^{i}}{C_{AS}} \right) + \left(\sum_{i=1}^{\text{min}(S,E)} \frac{C_{D}^{i}}{C_{MEC}} \right),
	\end{split}
	\label{computing_delay}
\end{equation}
where the preprocessing delay $t_n^{prep,S}$ is the time required to prepare the data for transmission to the MEC server, 
$t_n^{AV}(E,S) = t_n^{prep,S} + t_n^{proc}(E,S) = t_n^{prep,S} + \sum_{i=1}^{\min(S,E)} \frac{C_D^i}{C_{AS}}$ 
represents the local computing delay on the AV including the preparation of data for transfer to the MEC server and all CNN layers before the split point $S$ or the early exit $E$ (whichever occurs first), 
$t_n^{MEC}(E,S) = \sum_{i=S+1}^{E} \frac{C_D^i}{C_{MEC}}$ 
denotes the MEC computing delay on the MEC server, which accounts for the computing of all CNN layers that follow the selected split point $S$ up to the chosen exit $E$.

The total processing delay $t_n^{total}(E,S)$ is defined as the total time required to process the $n$-th CNN interference locally or offloaded to the MEC server and includes both the computing delay $t_n^{comp}(E,S)$ and the communication delay $t_n^{comm,S}$, i.e.,

\begin{equation}
	t^{total}_{n}(E,S) = t^{comp}_{n}(E,S) + t^{comm,S}_{n}.
	\label{total_processing_delay}
\end{equation}

Furthermore, we define the total energy consumption $E_n^{total}(E,S)$ of the AV for processing the $n$-th CNN interference as:

\begin{equation}
	\begin{split}
		&E_n^{total}(E,S) = E_n^{idle}(E,S) + E_n^{prep}(E,S) + \\
		& E_n^{comp}(E,S) + E_n^{comm}(E,S) = \\
		&= (t_n^{comm}(E,S) + t_n^{MEC}(E,S)) \times P_{idle} + \\
		&+ t_n^{prep,S} \times P_{prep} + t_n^{proc}(E,S) \times P_{comp} + \\
		& + t_n^{total}(E,S) \times P_{comm}.
	\end{split}
	\label{total_energy_consumption}
\end{equation}

\noindent where $E_n^{idle}(E,S)$  is the idle energy consumed by the AV’s CPU while waiting for results from the MEC server, $E_n^{prep,S}$, $E_n^{comp}(E,S)$, and $E_n^{comm}(E,S)$ represent the energy consumed for data preprocessing, local CNN computing, and communication, respectively. Each energy component is proportional to the duration of its corresponding phase and the associated power consumption $P_{idle}$, $P_{prep}$, $P_{proc}$, and $P_{comm}$.

%\section{Objectives of the work} \label{Problem}
%
%The objective of this paper is to evaluate the impact of the early exits and the splits in CNN architecture on the total processing delay $t_n^{total}$ and total energy consumption $E_n^{total}$. The selection of the early exit $E$ and the split point $S$ affects the total processing delay $t_n^{total}$ consisting of the computing delay $t_n^{comp}$ and the overall communication delay $t_n^{comm}$. Besides, selection of $S$ and $E$ influences also the total energy consumption $E_n^{total}$ composed of the computing energy $E_n^{comp}$, idle energy $E_n^{idle}$, preprocessing energy $E_n^{prep}$, and communication energy $E_n^{comm}$.

\section{Implementation of CNN to AV and mobile network with MEC} \label{Implementation}

This section describes the implementation of CNN with early exits and split points, enabling (partial) offloading to the MEC server, and the training process. Details of the physical AV architecture and MEC integration are available in \cite{Danek2025}.

\subsection{Architecture of CNN with early exit and split points}

%\textcolor{ForestGreen}{This subsection describes the implementation details of CNN with early exits and split points.} 
The implemented CNN, designed for road sign classification, is based on VGG-16 \cite{Gu2022}, leveraging its deep hierarchical structure and 3x3 convolutional kernels for efficient feature extraction in image classification tasks. However, its computational complexity limits real-time deployment in resource-constrained AVs. The implemented CNN consists of five blocks $(N_B=5)$, each followed by an exit of CNN, resulting in four early exits and one main exit (Fig.~\ref{Fig:neuralNetwork}). %\textcolor{ForestGreen}{Early exits allow faster classification at the cost of lower classification accuracy $A(E,S)$, while later exits achieve higher classification accuracy $A(E,S)$ but require more computation.}

Furthermore, six split points are introduced between the blocks to enable partial offloading. One additional split point before the first convolution allows full offloading. At each split point, computation on the AV can be terminated, and the extracted feature representations are transmitted to the MEC server for further processing. 

Since the total data volume at split points exceeds the input image size leading to higher communication delays in case of selection of later split points. To mitigate this, a compression mechanism based on autoencoders \cite{Matsubar2023} is applied. %\textcolor{ForestGreen}{Although the autoencoder slightly reduces classification accuracy $A(E,S)$ \cite{Schlett2023}, the impact remains within a few percent. }

%TODO \textcolor{ForestGreen}{
%\subsection{Training of CNN with early exit and split points}
%}

%TODO \textcolor{ForestGreen}{The implemented CNN is trained using the adaptive moment estimation (ADAM) optimizer \cite{Nikbakhtsarvestani2025} with a learning rate of 0.0001, chosen for its fast convergence and adaptive control of the learning step without complex parameter tuning. Cross-entropy \cite{Nikbakhtsarvestani2025} is employed as the loss function due to its proven effectiveness for multi-class classification.
%}
%TODO\textcolor{red}{tyto texty bych vsechny ponechal, ale treba deleni treninku do buletu zbytecne natahuje text, staci to nechat v textu a zkratit}
%TODO\textcolor{ForestGreen}{
%The training proceeds is divided into three phases \cite{Li2023}:
%}
%\textcolor{ForestGreen}{
%\begin{enumerate}[label=\roman*.] 
%	\item \uline{Training of the Whole CNN:} First, only the main exit of the CNN is trained (i.e., whole CNN is always processed). The CNN is trained to optimize classification accuracy at the final exit, while all early exits and compression layers are disabled.
%	\item \uline{Training of the Early Exits:} The CNN weights are frozen except for the layers forming early exits (highlighted in Fig.~\ref{Fig:neuralNetwork}), which are trained independently. 
%	\item \uline{Training of the Autoencoders:} Finally, only the autoencoder layers are trained to minimize classification loss, ensuring that compression does not significantly degrade classification accuracy. 
%\end{enumerate}
%}

The CNN is trained and validated on the German Traffic Sign Recognition Benchmark (GTSRB) dataset \cite{Stallkamp2012}, comprising 39,209 training and 12,630 validation images across 43 road sign classes.

\subsection{Integration of AV and MEC server}
The CNN can either be processed locally or (partially) offloaded to the MEC server over a software-defined 5G mobile network, implemented by means of OpenAirInterface\footnote{\href{https://openairinterface.org}{https://openairinterface.org}} (OAI)  and USRP N310 software-defined radio. Communication is handled via Internet protocol (IP) and transmission control protocol (TCP). The same CNN model is deployed on both the AV and the MEC server using Python and its library PyTorch. Data flow of CNN processing is shown in Fig~\ref{dataFlow}. For a detailed description of the AV’s architecture, hardware components, and software integration, please refer to \cite{Danek2025} or our GitLab\footnote{\href{https://gitlab.fel.cvut.cz/mobile-and-wireless/autonomous-driving/autonomous-driving-ROS1}{https://gitlab.fel.cvut.cz/mobile-and-wireless/autonomous-driving/autonomous-driving-ROS1}}.

%\begin{figure}[tb]
%	\centerline{\includegraphics[width=0.8\linewidth]{image/dataFlowSmall.png}}
%	\caption{Data flows from the AV’s camera to either (partial) offloading or full local processing on the AV. In the offloading scenario, data is transmitted via a mobile network to the MEC server for further computation. In every case, results from CNN are used in control algorithms of AV. }
%	\label{dataFlow}
%\end{figure}

\section{Experimental setup and scenario} \label{ExperimentalSetup}
%In this section, we discuss the setup and the scenario for experiments to evaluate benefits of the early exits and splits for CNN-based tasks offloading to MEC server. Measurement of performance metrics is also explained in this section.

This section outlines the experimental setup and scenario used to evaluate early exits and splits for CNN offloading to the MEC server and explain measured performance metrics.

\subsection{Experimental setup}
The MEC server uses a CPU with $C_{CPU}$ of 20.4 GFLOPS and a GPU with $C_{GPU}$  of 52000 GFLOPS corresponding to Intel Core i7-9700K CPU and NVIDIA RTX 2080 Ti GPU, respectively. The AV is equipped with computing power $C_{AV}$ of 7.36 GFLOPS provided by Intel Atom x7-E3950 CPU. 
Communication between the AV and the MEC server is established via 5G network at 3.5 GHz (n78) with 20 MHz bandwidth and 106 resource blocks, using modulation and coding scheme 24 (64QAM, code rate 0.754) \cite{3GPP2022}. This configuration is one of the commonly used setups in OAI with using USRP N310.

Table~\ref{tab:split_parameters_extended} presents parameters for each split point $S$ including, original data volume before compression $D_{orig}^S$, compressed data volume $D_{comp}^S$, transmitted data volume including TCP overhead and changes in coding due to Python $D_{ul}^S$, compression ratio $R_S$, returned result volume with TCP overhead $D_{dl}^S$ and without TCP overhead $D_{without}^S$, and computing demand $C^S_D$. 
%\textcolor{red}{nekonzistetni znaceni, dl index je ale ul neni, navic pro downlink je jednou index dl a jednou ul, pro duplink je pak i index dl} 
Note that autoencoders are applied for partial offloading (i.e., $S=1$–$4$).

\begin{table}[t]
	\vspace{1ex}
	\caption{Volume of data transferred in each split point $D_{ul}^S$ and $D_{dl}^S$ (including overhead) and computing demand for layers between split points $C_D^S$.}

	\centering
	\scriptsize
	\renewcommand{\arraystretch}{1.2}
	\setlength{\tabcolsep}{4pt}
	\resizebox{\columnwidth}{!}{
		\begin{tabular}{|c|c|c|c|c|c|c|c|}
			\hline
			\textbf{Split $S$} & \multicolumn{4}{c|}{\textbf{Uplink communication}} & \multicolumn{2}{c|}{\textbf{Downlink communication}} & \multicolumn{1}{c|}{\textbf{Computing demand}} \\
			\cline{2-8}
			& $D_{orig}^S$ [kb] & $D_{comp}^S$ [kb]  & $R_S$ [-] & $D_{ul}^S$ [kb]  & $D_{without}^S$ [kb] & $D_{dl}^S$ [kb] & $C_D^S$ [GFLOPS] \\
			\hline
			0 & 10.10 & 10.10 & 1 & 1749.8 & 1.4 & 1.6  & 0.145 \\
			\hline
			1 & 56.25 & 7.03  & 8 & 1206.4 & 1.4 & 1.6  & 0.226 \\
			\hline
			2 & 3.52  & 0.44  & 8 & 625.1  & 1.4 & 1.6  & 0.358 \\
			\hline
			3 & 1.53  & 0.19  & 8 & 279.4  & 1.4 & 1.6  & 0.311 \\
			\hline
			4 & 0.56  & 0.07  & 8  & 100.6 & 1.4 & 1.6  & 0.080 \\
			\hline
			5 & 0     & 0     & NaN   & 0   & 0   & 0      & 0  \\
			\hline
		\end{tabular}
	}
	\label{tab:split_parameters_extended}
\end{table}

\subsection{Performance metrics} \label{PerformanceMetrics}

We consider the following performance metrics (see Section II): i) total energy consumption defined in (\ref{total_energy_consumption}), ii) total processing delay, defined in (\ref{total_processing_delay}), and iii) classification accuracy, defined in  (\ref{classification_accuracy}). 

The communication energy $E_n^{comm}(E,S)$, measured by a USB power meter with 1 mWh resolution and ±1.41~\% accuracy, captures 5G modem consumption during uplink and downlink communication and during computation on MEC server. The computing $E_n^{comp}(E,S)$, idle $E_n^{idle}(E,S)$, and preprocessing $E_n^{prep}(E,S)$ energies are measured via Intel RAPL \cite{Intel2023}, which measures the energy consumed by the AV's processor in different phases of offloading.
%\textcolor{red}{takto to napsat nejde, je treba uvest presne co a jak presne se pouziva a pak dat referenci \cite{Intel2023}.}

To ensure precise measurement of uplink and downlink communication delays, the time bases of the AV and the MEC server are synchronized using the Network Time Protocol (NTP) \cite{Tripathi2021}.

The classification accuracy $A(E,S)$ is obtained by processing all images through specific early exit $E$ and split point $S$.

\subsection{Scenario of experiment}

%To ensure consistent testing conditions, the mobile network data rate is limited to 25~Mbps in both uplink and downlink directions. Road sign images used in the experiments are randomly selected from the GTSRB dataset.
%
%Each experiment processes a single road sign image through the CNN, either locally or with (partial) offloading, while recording total processing delay $t_n^{total}(E,S)$, total energy consumption $E_n^{total}(E,S)$, and classification accuracy $A(E,S)$. Each experiment is repeated 250 times, and results are averaged.

To ensure consistent conditions, the mobile network data rate is limited to 25~Mbps in both uplink and downlink. Road sign images are randomly selected from the GTSRB dataset.

Each experiment processes single road sign image through the CNN, either locally or with (partial) offloading, while recording total processing delay $t_n^{total}(E,S)$, energy consumption $E_n^{total}(E,S)$, and classification accuracy $A(E,S)$. Each experiment is repeated 250 times and results are averaged.

%\begin{figure}[tb]
%	\centerline{\includegraphics[width=0.8\linewidth]{image/map.png}}
%	\caption{Illustrative map of environment for experiments showing space for AV movement (grey area), walls (black lines), obstacles (red boxes), example of computed path for AV (green line), AV’s start and destination, base station (80 cm above gorund), MEC server, and road signs.}
%	\label{experimentalSetup}
%\end{figure}

\section{Performance evaluation} \label{Evaluation}
In this section, we evaluate offloading of the CNN with splits and early exits.  We present the total processing delay $t_n^{total}(E,S)$, total energy consumption $E_n^{total}(E,S)$, and classification accuracy $A(E,S)$ as functions of the split point $S$ and early exit $E$. 
Note that the full offloading corresponds to $S=0$ while full local processing to $S=5$. If $E \leq S$, the CNN is processed entirely on the AV; otherwise, computation is split between the AV and MEC server.

%Fig~\ref{classificationAccuracy} shows the classification accuracy $A(E,S)$ as a function of early exit $E$ and split point $S$. The classification accuracy $A(E,S)$ increases with later exits from 32~\% (early exit 1) to 93~\% (main exit), confirming that processing of more CNN layers enhances accuracy as expected. The most significant improvement (from about 30\% to more than 80\%) in classification accuracy $A(E,S)$ is observed between the first three exits, while the classification accuracy $A(E,S)$ gain between early exit 3 and the main exit is only several percent. The classification accuracy $A(E,S)$ remains almost constant disregarding split points for each early exit indicating that the compression by the autoencoders has only negligible impact. 

\begin{figure*}[t]
	\centering
	% Horní řada
	\begin{minipage}{0.32\linewidth}
		\centering
		\includegraphics[width=\linewidth]{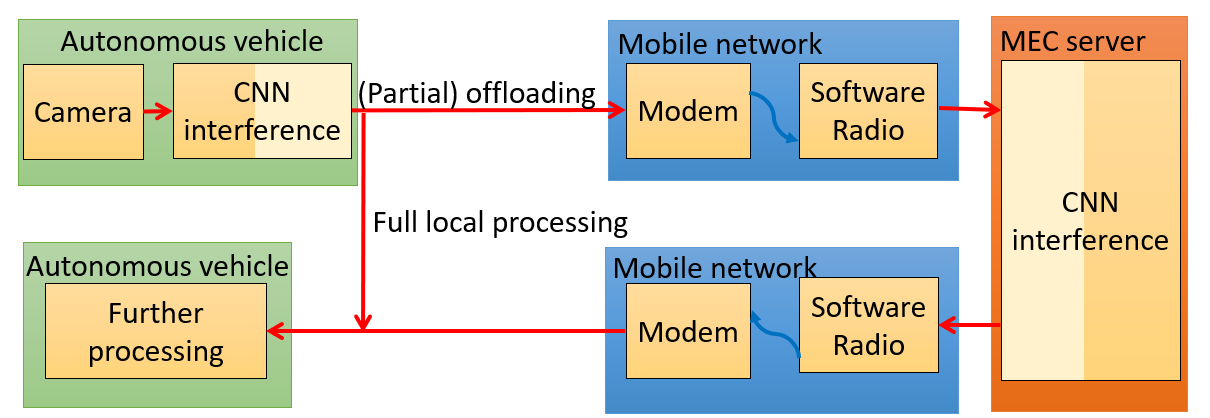}
		\caption{
			Data flows from the AV’s camera to either (partial) offloading or full local processing on the AV. In the offloading scenario, data is transmitted via a mobile network to the MEC server for further computation. 
			}
		\label{dataFlow}
	\end{minipage}\hfill
	\begin{minipage}{0.32\linewidth}
		\centering
		\includegraphics[width=0.85\linewidth]{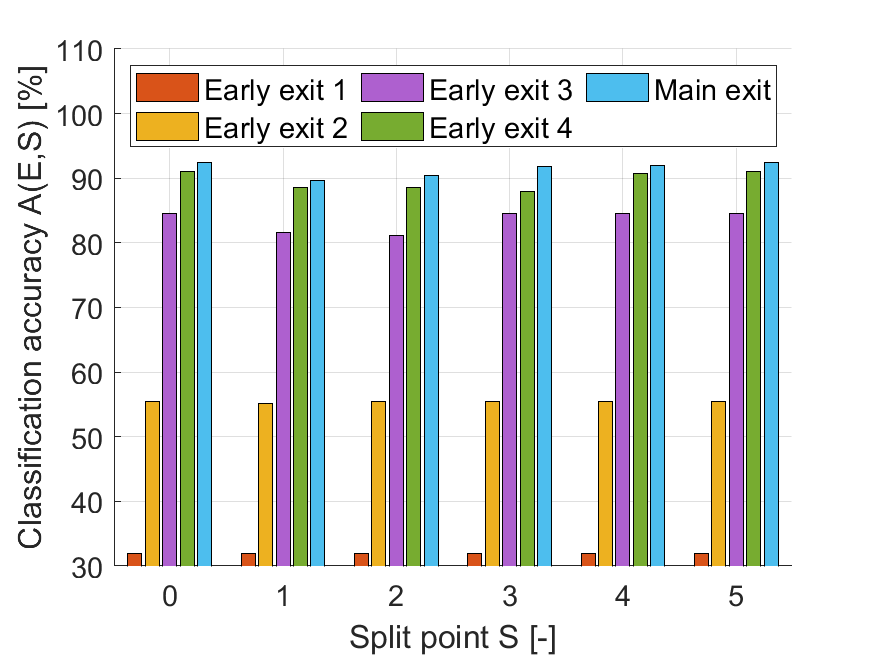}
		\caption{Impact of split point $S$ and early exit $E$ on classification accuracy $A(E,S)$.  }
		\label{classificationAccuracy}
	\end{minipage}\hfill
	\begin{minipage}{0.32\linewidth}
		\centering
		\includegraphics[width=0.85\linewidth]{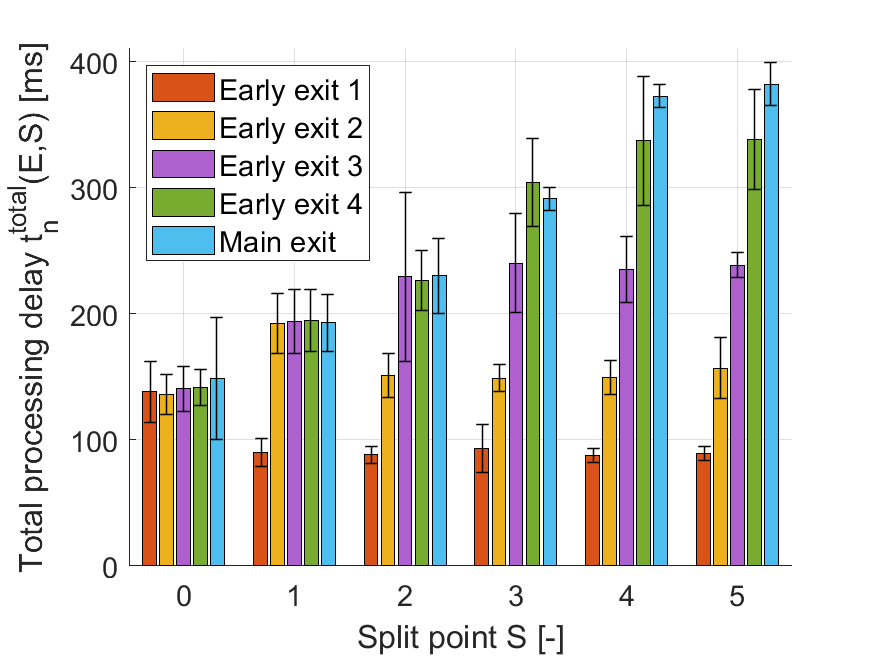}
		\caption{Impact of split point $S$ and early exit $E$ on total processing delay $t_n^{total}(E,S)$. Whiskers represent a standard deviation (68\% confidence).}
		\label{time}
	\end{minipage}\hfill
	
	% Spodní řada
	\begin{minipage}{0.32\linewidth}
		\centering
		\includegraphics[width=0.85\linewidth]{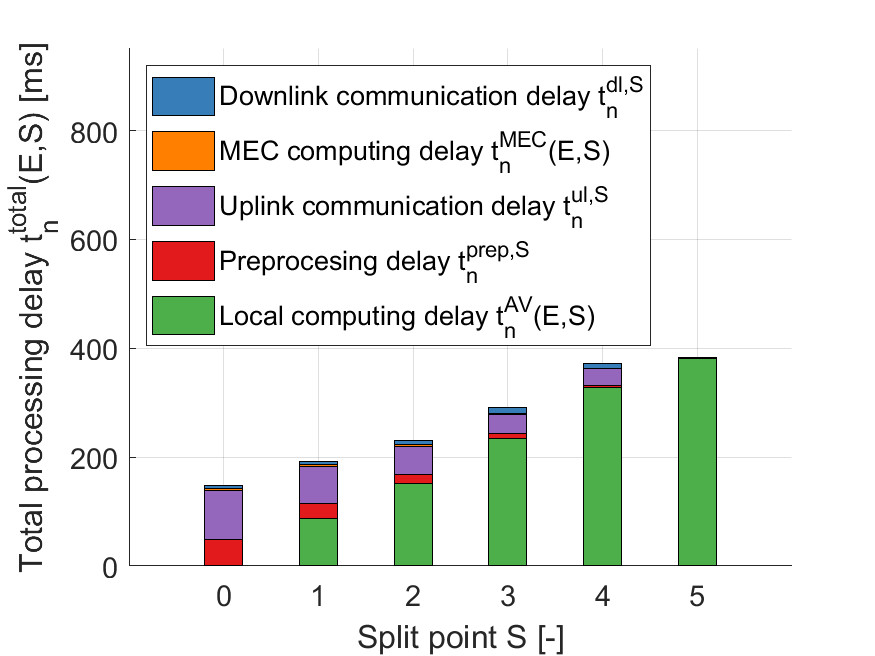}
		\caption{Impact of split $S$ on communication, computing, and preprocessing delays for main exit $(E=5)$.}
		\label{timeDetail}
	\end{minipage}
	\begin{minipage}{0.32\linewidth}
		\centering
		\includegraphics[width=0.85\linewidth]{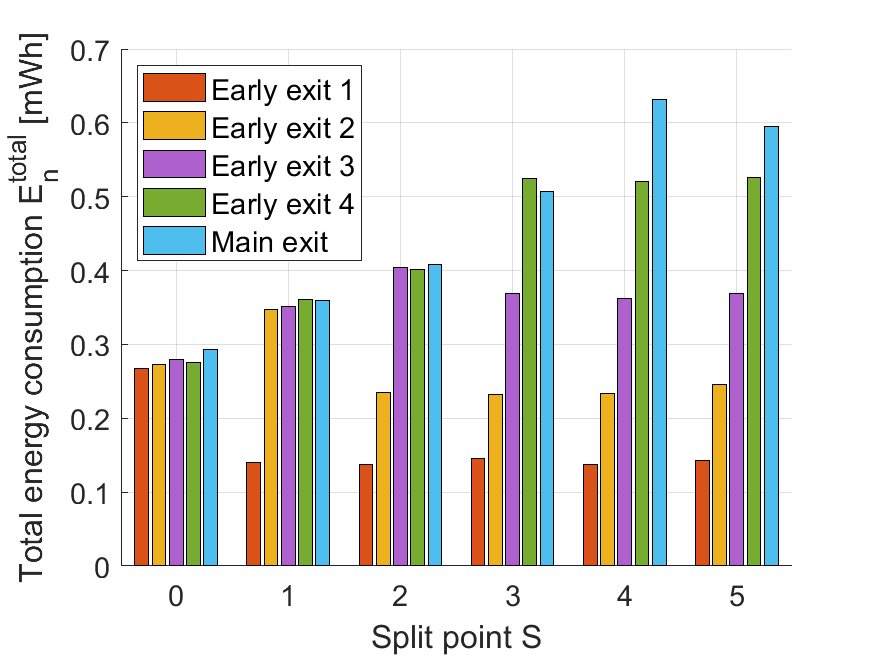}
		\caption{Impact of splits point $S$ and early exits $E$ on total energy consumption $E_n^{total}(E,S)$. }
		\label{energy}
	\end{minipage}\hfill
	\begin{minipage}{0.32\linewidth}
		\centering
		\includegraphics[width=0.85\linewidth]{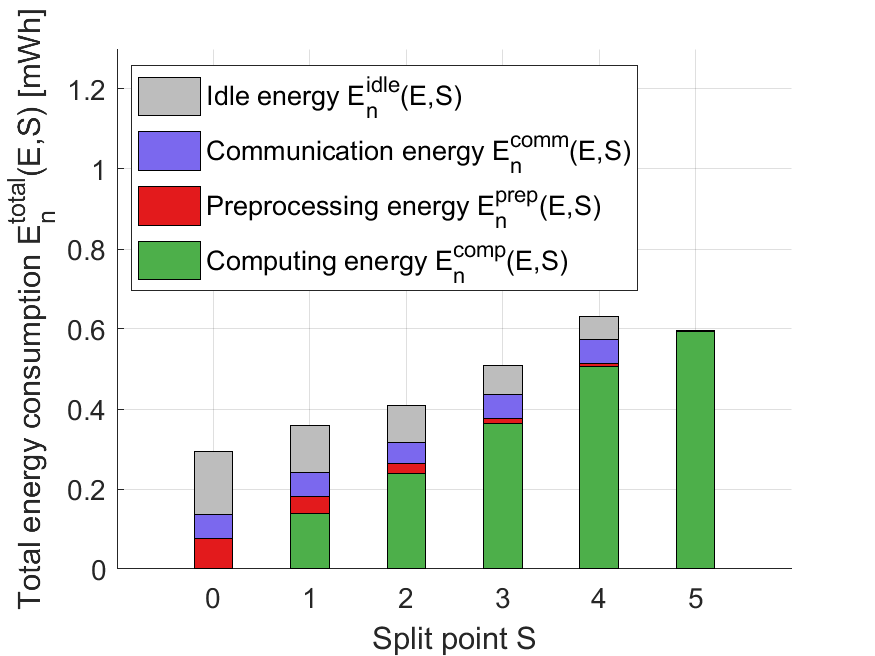}
		\caption{Impact of split $S$ on communication, computing, and preprocessing energy consumption for main exit $(E=5)$.}
		\label{energyDetail}
	\end{minipage}
\end{figure*}

Figure~\ref{classificationAccuracy} shows the classification accuracy $A(E,S)$ depending on early exit $E$ and split point $S$. The  $A(E,S)$ rises from 32\% (exit~1) to 93\% (main exit), confirming that later  CNN layers improve $A(E,S)$. The most notable gain (from about 30\% to more than 80\%)  occurs between the first three exits and further increase beyond exit~3 is only marginal. For each exit, classification accuracy $A(E,S)$ stays nearly constant across all split points, indicating minimal impact of autoencoder compression.

%\begin{figure}[tb]
%	\centerline{\includegraphics[width=0.8\linewidth]{image/dataFlowSmall.png}}
%	\caption{Data flows from the AV’s camera to either (partial) offloading or full local processing on the AV. In the offloading scenario, data is transmitted via a mobile network to the MEC server for further computation. In every case, results from CNN are used in control algorithms of AV. }
%	\label{dataFlow}
%\end{figure}

%Fig.~\ref{time} shows the total processing delay $t_n^{total}(E,S)$  for different split points $S$ and exits $E$. If the split point is located beyond the early exit, all preceding layers are processed locally. The results demonstrate that selecting earlier exits significantly reduces total processing delay $t_n^{total}(E,S)$, as fewer CNN layers need to be computed. Conversely, increasing the split $S$ keeps more computation at the AV, which leads to higher local computing delay $t_n^{AV}(E,S)$ due to AV limited computing power $C_{AV}$. 
%The early exits significantly (up to 4.2 times) reduce the total processing delay compared to the main exit. Total processing delay increases with later split points as more layers are computed on the less computationally powerful AV before offloading to the powerful MEC, as expected according to (\ref{communication_delay}). 

Figure~\ref{time} shows the total processing delay $t_n^{total}(E,S)$ for various exits $E$ and split points $S$. 
%TODO If the split point is located beyond the early exit, all preceding layers are processed locally. 
Earlier exits reduce $t_n^{total}(E,S)$ significantly by limiting the number of computed layers. Conversely, later splits increase the local computing delay $t_n^{AV}(E,S)$ due to AV's limited computing power $C_{AV}$. Overall, early exits reduce $t_n^{total}(E,S)$ by up to 4.2 times compared to the main exit, with later splits increasing $t_n^{total}(E,S)$ as more layers are computed on the less computationally powerful AV before offloading, as expected according to (\ref{computing_delay}). Full offloading reduces the $t_n^{total}(E,S)$ by up to 2.5 times compared to full local processing on the AV.

%Fig.~\ref{timeDetail} shows that the uplink communication delay $t_n^{ul,S}$ decreases with increasing split points, in line with (\ref{communication_delay}), due to a lower volume of data $D_n^{ul,S}$ to be transferred, see Table~\ref{tab:split_parameters_extended}. Since the split has no impact on downlink data volume, the downlink communication delay $t_n^{dl,S}$ remains constant except for $S=5$, where no offloading occurs.
%The local computing delay $t_n^{AV}(E,S)$ increases, as the position of the split moves toward later stages of the CNN, requiring computation of more layers at the AV. In contrast, the MEC computing delay $t_n^{MEC}(E,S)$  decreases for higher splits. However, the MEC computing delay is up to 81.5 times lower than the local computing delay $t_n^{AV}(E,S)$, reflecting the higher computing power of the MEC server $C_{MEC}(E,S)$.
%The preprocessing delay $t_n^{prep}(E,S)$ is considered only for the uplink, as $t_n^{prep,S}$ varies with the volume of data transmitted from the AV to the MEC server $D_{ul}^S$, while the preparation of single classification result on the MEC server is negligible and not measurable. 

Figure~\ref{timeDetail} shows that uplink communication delay $t_n^{ul,S}$ decreases with higher split points, as the transmitted data volume $D_ul^{S}$ shrinks between splits, see Table~\ref{tab:split_parameters_extended}, in line with (\ref{communication_delay}). Downlink communication delay $t_n^{dl,S}$ remains constant, except at $S{=}5$, where no offloading occurs. 
As the split moves moves toward later stages of the CNN, local computing delay $t_n^{AV}(E,S)$ increases due to more layers being computed at the AV, while MEC computing delay $t_n^{MEC}(E,S)$ decreases. However, MEC computing delay is up to 81.5 times lower than the local computing delay $t_n^{AV}(E,S)$, reflecting higher computing power $C_{MEC}$ of MEC server. 
The preprocessing delay $t_n^{prep,S}$ is considered only for the uplink, as $t_n^{prep,S}$ varies with the volume of data transmitted from the AV to the MEC server $D_{ul}^S$, while the preparation of single classification result on the MEC server is negligible and not measurable. 

%Fig.~\ref{energy} shows the total energy consumption $E_n^{total}(E,S)$ as a function of splits and early exits, illustrating the trade-offs between local computing and offloading. The early exits significantly reduce the energy consumption (up to 4.4 times) compared to the main exit.  Full offloading ($S=0$) keeps $E_n^{total}(E,S)$ low across all exits, as the AV's CPU performs only light computation and remains mostly idle while the more powerful MEC server handles the CNN processing. Additionally, the fast computing on the MEC server shortens the total processing time $t_n^{total}(E,S)$, which further reduces the total energy consumption. The higher split points shift more computation to the AV, thus, leading to increasing energy consumption.

Figure~\ref{energy} shows the total energy consumption $E_n^{total}(E,S)$ across exits and split points, highlighting the trade-off between local computing and offloading. Early exits reduce energy consumption by up to 4.4 times compared to the main exit. Full offloading ($S=0$) keeps $E_n^{total}(E,S)$ low, as the AV remains mostly idle while the MEC server handles CNN inference. Fast MEC computation also shortens $t_n^{total}(E,S)$, further reducing energy. In contrast, higher splits shift more computation to the AV, increasing energy consumption.

%Fig.~\ref{energyDetail} further breaks down the energy consumption for the main exit into individual components. For the full local processing ($S=5$), the communication energy is vanished, since no communication takes place. At the same time, the total energy is 2.6 times higher compared to full offloading, due to significantly more energy  $E_n^{AV}(E,S)$ consumed for full CNN processing by the AV. For the full offloading ($S=0$), the total energy consumption is dominated by idle energy $E_n^{idle}(E,S)$, as the AV waits for the MEC server to complete the CNN's processing. In this case, the computing energy at the AV  $E_n^{AV}(E,S)$ is zero. As the split point increases, a larger portion of CNN is processed locally on the AV, resulting in higher computing energy $E_n^{AV}(E,S)$.

Figure~\ref{energyDetail} breaks down the energy consumption for the main exit into individual components. With full local processing ($S=5$), communication energy $E_n^{comm}(E,S)$ is zero, but total energy consumption $E_n^{total}(E,S)$ is 2.6 times higher than with full offloading due to increased $E_n^{comp}(E,S)$. In contrast, full offloading ($S=0$) results in zero computing energy at the AV, with total energy consumption $E_n^{total}(E,S)$ dominated by idle energy $E_n^{idle}(E,S)$, as the AV waits for the MEC server to complete the CNN's processing. As the split increases, more CNN layers are computed locally, raising $E_n^{comp}(E,S)$.

\section{Modeling of total processing time and total energy consumption} \label{Modeling}

The experimental results presented in Section~\ref{Evaluation}, allow us to create realistic models of the total processing time $t_n^{total}(E,S)$ and the total energy consumption $E_n^{total}(E,S)$.  The total processing delay is theoretically defined in (\ref{total_processing_delay}), however, the following expressions for $t_n^{proc}(E,S)$, $t_n^{MEC}(E,S)$, and $t_n^{prep,S}$ provide refined formulations reflecting the real-world conditions observed during the experiments: 

\begin{equation}
	t_n^{proc}(E,S) = \sum_{i=1}^{\min(S,E)}  \left( d_{AV} + \frac{C_D^i}{C_{AS}}\right),
	\label{local_processing_time_modeling}
\end{equation}

\begin{equation}
	t_n^{MEC}(E,S) = \sum_{i=S+1}^{E} \left( d_{MEC} + \frac{C_D^i}{C_{MEC}} \right),
	\label{MEC_processing_time_modeling}
\end{equation}

\begin{equation}
	t_n^{prep,S} = d_{prep} +k_{prep} D^S_{comp},
	\label{preprocessing_time_modeling}
\end{equation}
where $d_{AV}$  and $d_{MEC}$ are the constant processing delays on the AV and MEC server, $d_{prep}$ is the fixed delay at the start of preprocessing, while $k_{prep}$ captures the delay that increases proportionally with the volume of transmitted data $D^S_{comp}$. Estimated parameter values are summarized in  Table~\ref{processing_energy_parameters}.

In contrast to reformulated expressions above, the expression for $t_n^{comm,S}$  in (\ref{communication_delay}) matches real-world experiment.

The total energy consumption $E_n^{total}(E,S)$ is modeled based on (\ref{total_energy_consumption}) and the estimated power consumption values  $P_{idle}$, $P_{prep}$, $P_{proc}$, and $P_{comm}$ are provided in Table~\ref{processing_energy_parameters}.

The estimated values for models, we present values for all variables in Table~\ref{processing_energy_parameters}. The higher computing power of the MEC server relative to the AV explains the reduction in inference time upon offloading. The communication delay parameters match the 5G network configuration described in section~\ref{ExperimentalSetup}. 

\begin{table}[t]
	\vspace{0.7ex}
	\caption{Values of parameters for estimating total processing delay $t_n^{total}$ and total energy consumption $E_n^{total}$.}
	\centering
	\scriptsize
	\renewcommand{\arraystretch}{1.2}
	\setlength{\tabcolsep}{3pt}
	\resizebox{\columnwidth}{!}{
		\begin{tabular}{|c|c||c|c||c|c|}
			\hline
			\textbf{Parameter} & \textbf{Value} & \textbf{Parameter} & \textbf{Value} & \textbf{Parameter} & \textbf{Value} \\
			\hline
			$d_{ul}$ [ms] & 22.81 & $b_{ul}$ [Mbps] & 12.36 & $d_{dl}$ [ms] & 7.19 \\
			\hline
			$b_{dl}$ [Mbps] & 9.81 & $d_{AS}$ [ms] & 43.69 & $d_{MEC}$ [ms] & 1.12 \\
			\hline
			$d_{prep}$ [ms] & 12.18 & $k_{prep}$ [kb/ms] & 2.33 & $C_{MEC}$ [GFLOPS] & 365.94 \\
			\hline
			$C_{AS}$ [GFLOPS] & 3.62 & $P_{idle}$ [W] & 4.62 & $P_{prep}$ [W] & 4.92 \\
			\hline
			$P_{proc}$ [W] & 5.17 & $P_{comm}$ [W] & 0.79 & & \\
			\hline
		\end{tabular}
	}
	\label{processing_energy_parameters}
\end{table}

\section{Conclusion} \label{Conclusion}

We have demonstrated the feasibility of computation offloading from autonomous systems to the MEC server using a real AV and software define mobile network infrastructure. Focusing on road sign classification, we have evaluated CNN enhanced with early exits and split points to support partial or full offloading. Our results show that computation offloading reduces the total processing delay by up 2.5 times and the total energy consumption by up to 2.6 times compared to the local processing by the AV.
The classification accuracy is influenced by the choice of the split point and the early exit. The exits located after more CNN layers improve classification accuracy at the cost of higher computing delay. We also developed analytical models for the total processing delay and the total energy consumption based on the real-world experiments.

Future work should expand the system to multiple BS and AVs and explore new offloading decision algorithm for choosing suitable split and early exit on fly according to actual channel conditions and resource availability.


\begin{thebibliography}{00}
	\bibitem{Karangwa2023} J. Karangwa, J. Liu and Z. Zeng, "Vehicle Detection for Autonomous Driving: A Review of Algorithms and Datasets," IEEE T-ITS, 2023.
%	\bibitem{Becvar2017}  Z. Bečvář and P. Mach, "Mobile Edge Computing: A Survey on Architecture and Computation Offloading," IEEE Communications Surveys \& Tutorials, vol. 19, no. 3, 2017.
	\bibitem{Liu2022} L. Liu, et al. , "Deep Reinforcement Learning-based Dynamic SFC Deployment in IoT-MEC Networks," IEEE ICCC, 2022. 
%	\bibitem{Li2022} X. Li, S. Bi, Z. Quan and H. Wang, "Online Cognitive Data Sensing and Processing Optimization in Energy-Harvesting Edge Computing Systems," IEEE Transactions on Wireless Communications, 2022. 
	\bibitem{Hu2022} H. Hu, et al., "Energy Efficiency and Delay Tradeoff in an MEC-Enabled Mobile IoT Network," IEEE Internet of Things Journal, 2022. 
%	\bibitem{Sheng2020} M. Sheng, et al., "Delay-Aware Computation Offloading in NOMA MEC Under Differentiated Uploading Delay," IEEE Transactions on Wireless Communications, 2020. 
	\bibitem{Danek2025} J. Danek, Z. Becvar and A. Janes, "Computational Offloading for Autonomous Systems: Real-World Experiments and Modeling," IEEE Vehicular Technology Conference (IEEE VTC2025-Spring), 2025. 
	\bibitem{LiN2022} N. Li, et al., Graph Reinforcement Learning-based CNN Inference Offloading in Dynamic Edge Computing. IEEE GLOBECOM, 2022.
	\bibitem{Casale2024} G. Casale and M. Roveri, "Scheduling Inputs in Early Exit Neural Networks," in IEEE Transactions on Computers, vol. 73, no. 2, 2024.
	\bibitem{Teerapittayanon216} S. Teerapittayanon, et al., "Branchynet: Fast inference via early exiting from deep neural networks," IEEE ICPR, 2016.
	\bibitem{Bakhtiarnia2024} A. Bakhtiarnia, et al., "Dynamic Split Computing for Efficient Deep EDGE Intelligence," IEEE ICASSP, pp. 1-5, 2024. 
%	\bibitem{Huang2024} B. Huang,, L. Pang, A. Fu, S. F. Al-Sarawi and D. Abbo, "Sponge Attack Against Multi-Exit Networks With Data Poisoning," IEEE Access, 2024. 
%	\bibitem{Wang2024} J. Wang, B. Li and G. L. Zhang, "Early-Exit with Class Exclusion for Efficient Inference of Neural Networks," IEEE AICAS, 2024.
	\bibitem{Zhou2023} H. Zhou, et al., "Accelerating Deep Learning Inference via Model Parallelism and Partial Computation Offloading," IEEE TPDS, 2023.
	\bibitem{Eshratifar2021} A. E. Eshratifar, M. S. Abrishami and M. Pedram, "JointDNN: An Efficient Training and Inference Engine for Intelligent Mobile Cloud Computing Services," IEEE Trans. Mob. Comput., 2021.
%	\bibitem{Tuli2023} S. Tuli, G. Casale and N. R. Jennings, "SplitPlace: AI Augmented Splitting and Placement of Large-Scale Neural Networks in Mobile Edge Environments," IEEE Transactions on Mobile Computing, 2023.
%	\bibitem{Bajpai2024} D. J. Bajpai, A. Jaiswal and M. K. Hanawal, "I-SplitEE: Image Classification in Split Computing DNNs with Early Exits," ICC 2024 - IEEE International Conference on Communications, pp. 2658-2663, 2024.
%	\bibitem{Pacheco2024} R. G. Pacheco, . F. D. V. R. Oliveira and R. S. Cout, "Early-exit deep neural networks for distorted images: providing an efficient edge offloading," IEEE GLOBECOM, 2021. 
	\bibitem{Narmeen2024} R. Narmeen, P. Mach, Z. Becvar and I. Ahmad, "Joint Exit Selection and Offloading Decision for Applications Based on Deep Neural Networks," IEEE Internet of Things Journal, vol. 11, no. 23, pp. 38098-38112, 2024. 
	\bibitem{Rauch2024} R. Rauch, Z. Becvar, P. Mach and J. Gazda, "Cooperative Multi-Agent Deep Reinforcement Learning for Dynamic Task Execution and Resource Allocation in Vehicular Edge Computing," IEEE TVT, 2024.
	\bibitem{Matsubara2020} Y. Matsubara, et al., "Head Network Distillation: Splitting Distilled Deep Neural Networks for Resource-Constrained Edge Computing Systems," IEEE Access, 2020.
	\bibitem{Matsubar2023} Y. Matsubar, et al., "SC2 benchmark: Supervised compression for split computing," Transactions on machine learning research, 2023. 
	\bibitem{Xu2003} R. Xu, et al., "Impact of data compression on energy consumption of wireless-networked handheld devices," ICDCS, 2003.
	\bibitem{3GPP2022} 3GPP, "5G Mobile System Architecture, TS 23.501, version 17.5.0," PP Technical Specification, 2022.
%	\bibitem{Coll-Perales2023} B. Coll-Perales and et al., "End-to-End V2X Latency Modeling and Analysis in 5G Networks," IEEE TVT, 2023.
	\bibitem{Gu2022} X. Gu, Q. Lang, F. Lin and P. Wang, "Attention-aware CNN model for Traffic Signs Classification," CBASE, pp. 83-87, 2022.
%	\bibitem{Schlett2023} T. Schlett, S. Schachner, C. Rathgeb, J. Tapia and C. Busch, "Effect of Lossy Compression Algorithms on Face Image Quality and Recognition," IEEE ICASSP, pp. 1-5, 2023.
%	\bibitem{Saputro2022} P. H. Saputro, et al., "Comparison ADAM-optimizer and SGDM for Classification Images of Rice Leaf Disease," ICIMCIS, 2022. 
	\bibitem {Nikbakhtsarvestani2025} F. Nikbakhtsarvestani, S. Rahnamayan and M. Ebrahimi, "Training Deep Neural Networks with Multi-Objective Adam Optimizer for Medical Image Classification," IEEE CIHM, 2025.
%	\bibitem{Shin2023} J. Shin and W. Chung, "Multi-Band CNN With Band-Dependent Kernels and Amalgamated Cross Entropy Loss for Motor Imagery Classification," IEEE Journal of Biomedical and Health Informatics, 2023.
	\bibitem{Li2023} S. Li, H.-T. Nguyen and C. C. Cheah, "A Theoretical Framework for End-to-End Learning of Deep Neural Networks With Applications to Robotics," IEEE Access, vol. 11, pp. 21992-22006, 2023.
	\bibitem{Stallkamp2012} J. Stallkamp, M. Schlipsing, J. Salmen and C. Igel, "Man vs. computer: Benchmarking machine learning algorithms for traffic sign recognition," Neural Networks, 2012.
%	\bibitem{Kaltenberger2020} F. Kaltenberger, A. P. Silva, A. Gosain, L. Wang and T.-T. Nguyen, "Openairinterface: Democratizing innovation in the 5g era," Computer Networks, vol. 176, 2020.
	\bibitem{Intel2023}Intel Corporatio, "Running Average Power Limit (RAPL) Energy Reporting," 2023. [Online]. Available: ttps://www.intel.com/content/www/us/en/developer/articles/technical/
	software-security-guidance/advisory-guidance/running-average-power-limit-energy-reporting.htm. [Accessed 24 03 2025].
	\bibitem{Tripathi2021}N. Tripathi and N. Hubballi, "Preventing time synchronization in NTP broadcast mode," Computers \& Security, vol. 102, 2021.
%	\bibitem{Baziyad2021} M. Baziyad, et al., "Addressing Real-Time Demands for Robotic Path Planning Systems: A Routing Protocol Approach," IEEE Access, 2021.
	
	
\end{thebibliography}
\end{document}